\documentclass{article}

\usepackage{arxiv}

\usepackage[utf8]{inputenc} 
\usepackage[T1]{fontenc}    
\usepackage{hyperref}       
\usepackage{url}            
\usepackage{booktabs}       
\usepackage{amsfonts}       
\usepackage{nicefrac}       
\usepackage{microtype}      
\usepackage{lipsum}
\usepackage{graphicx}
\usepackage{natbib}
\usepackage{amsmath}
\usepackage[toc,page]{appendix}
\graphicspath{ {./images/} }

\title{Emergence of chaos in the tropical atmosphere: Study of the weak temperature gradient system}

\author{
 Stéphane Vannitsem \\
 School of Physical and Mathematical Sciences \& The Asian School of the Environment\\
 Nanyang Technological University\\
 Singapore\\
  \texttt{Stephane.Vannitsem@ntu.edu.sg} \\
  \And
 Jonathan Demaeyer\\
  Meteorological and Climatological Research Service\\
  Royal Meteorological Institute of Belgium\\
  Belgium \\
  \texttt{Jonathan.Demaeyer@meteo.be} \\
}

\begin{document}
\maketitle
\begin{abstract}
The atmospheric tropical belt is believed to be more predictable than the extratropics. This question is revisited here by exploring the emergence of chaos in reduced-order model versions of the vorticity equation under the weak temperature gradient hypothesis, which provides a good description of the large-scale tropical atmosphere. The analysis reveals that under fairly realistic divergence forcing amplitudes, chaos may emerge, sometimes with Lyapunov time scales of less than a day. This result contrasts with the idea of a predictable tropical atmosphere, and opens important questions on the effective origin of predictability in the Tropics.
\end{abstract}


\section{Introduction}
The question of the limit of predictability of the atmosphere traces back to the early work of \cite{Thompson1957} in which the gradual degradation of the forecasts due to the initial state uncertainty was emphasized. This question was then taken up in the celebrated paper by E.N. Lorenz showing the exponential amplifications of small errors in a simple 3-variable deterministic system \citep{Lorenz1963}. After these seminal works, the interest for the analysis of the predictability of the atmosphere either from a practical or theoretical point of view has exploded, with the analysis of a wide range of atmospheric and climate models \citep[e.g.][]{Kalnay2003, Boer2004} and the development of the theory of Chaos \citep[e.g.][]{Li1975, Ott2002}. 

In general, the main focus was to evaluate the predictability of the global atmosphere and its climate \citep{Kalnay2003} or with a focus on the extratropical regions in which baroclinicity is believed to be the main source of instability \citep[e.g.][]{Buizza1995,Vannitsem1997,Li2011}. An interesting work along this line shows that in the tropical regions, the uncertainty at the interface between the ocean and the atmosphere is the main source of limited predictability of the atmosphere and not its initial state \citep{Shukla1998}. The question is therefore to know whether the tropical atmosphere is essentially slaved to its boundary conditions. This question is taken up in the present paper by exploring the emergence of chaos in simplified equations describing the large scale dynamics of the atmosphere in the tropical regions.    

The tropical atmosphere has been demonstrated to be the place of large scale wave dynamics \citep{Matsuno1966,Gill1980,Zhang1996,Wheeler1999}. Since these seminal works mostly based on linear assumptions, considerable extensions have been made by incorporating additional processes, nonlinearities and stochastic forcing emulating fast processes \citep{Zhang1996,Kraucunas2007, Kiladis2009,Krishnamurti2013,Khouider2019, Yano2025}. But to the knowledge of the authors, there is no specific analysis of the emergence of chaos in the large-scale tropical atmosphere, a path we will taken up here.

To this aim, we will consider one of the simplest equations to represent the large scale dynamics of a dry tropical atmosphere based on the Weak Temperature Gradient (WTG) approximation \citep{Sobel2000, Bretherton2002, Bretherton2003, Emanuel2007, Smith2015, vallis2019, Adames2022}. In its simplest form, the system is limited to a unique nonlinear partial differential equation for the vorticity, while the divergence is considered as a constant source term. A reduced-order model will then be extracted leading to a finite set of ordinary differential equations. These new equations can then be analyzed through the computation of the Lyapunov exponents allowing to characterize the emergence of chaos as in the recent works on the extratropical multiscale dynamics \citep{Vannitsem2015, Vannitsem2017, Hamilton2023, Xavier2024}. 

Section 2 is devoted to the description of the WTG equation and the reduced-order model. The nature of the solutions emerging from these equations are then explored in Section 3 by computing the Lyapunov exponents characterizing the sensitivity to initial conditions. Future research lines are then drawn in Section 4.

\section{The reduced-order WTG model}

In this section, we first briefly review the basic equations from which the reduced-order model is deduced. The domain and the forcing term is then presented, together with the reduced-order model which is detailed in Appendix A.

\subsection{The basic equations}

The Weak Temperature Gradient (WTG) hypothesis \citep{Sobel2000,Sobel2001,Bretherton2002,Bretherton2003} assumes that the temperature is horizontally rapidly homogenized in the Tropics through the fast propagation of gravity waves. This has been used as a key hypothesis for the description of the large scale dynamics of the tropical atmosphere \citep[e.g.][]{Smith2015, vallis2019, Adames2022}. Under this assumption, the vertical velocity (here in pressure coordinates) is expressed as
\begin{equation}
  \omega = \frac{Q}{\sigma_p}  
\end{equation}
where $Q$ is the heat injected in the system and $\sigma_p$, the static stability. Introducing it in the continuity equation, we get the horizontal velocity divergence
\begin{equation}
  D  = \frac{\partial u}{\partial x}+ \frac{\partial v}{\partial y} = -\frac{\partial \omega}{\partial p}
\end{equation}
If the horizontal divergence $D=-\frac{\partial \omega}{\partial p}$ is assumed constant then the large-scale dry dynamics is governed by the vorticity equation,
\begin{eqnarray}
    \frac{\partial \xi}{\partial t} &= &- (\vec{v}_p \cdot \vec{\nabla}) (\xi +f) -(\xi +f)D - r \xi \label{eq:vorticity_PDE} \\
    {\nabla}^2_p \phi & = & - {\nabla}^2_p (\vec{v}_p^2 /2) +  \Bigl( v \frac{\partial \xi}{\partial x} - u \frac{\partial (\xi+f) }{\partial y} \Bigr) + \xi (\xi+f) \label{eq:div_PDE}
\end{eqnarray}
in which a linear dissipation has been added, controlled by the parameter $r$. The model is defined on a beta plane with zonal and meridional coordinates $x$ and $y$, and with the planetary component of the vorticity being defined as $f = f_0 + \beta y$. The second equation is a diagnostic equation for the geopotential. Note that as the divergence is non-zero, then an ageostrophic wind is present. The flow velocity, the divergence and the relative vorticity are defined as
\begin{eqnarray}
    u & = & -\frac{\partial \psi}{\partial y} + \frac{\partial \chi}{\partial x} \\
    v & = & \frac{\partial \psi}{\partial x} + \frac{\partial \chi}{\partial y}  \\
    D & = & {\nabla}^2_p \chi \\
    \xi & = & \nabla^2_p \psi
\end{eqnarray}
where $\psi$ is the streamfunction and $\chi$ the velocity potential.

The description of the dynamics then reduces to a unique prognostic equation,
\begin{equation}
    \label{eq:WTG_PDE}
    \frac{\partial}{\partial t}  \nabla^2 \psi = - J(\psi, \nabla^2 \psi) - \beta \frac{\partial \psi}{\partial x} - \nabla \chi \cdot \nabla (\nabla^2 \psi + f) -  (\nabla^2 \psi + f) \nabla^2 \chi - r \nabla^2 \psi
\end{equation}
where the Jacobian $J(\psi, \eta) = \frac{\partial \psi}{\partial x}\frac{\partial \eta}{\partial y} - \frac{\partial \psi}{\partial y}\frac{\partial \eta}{\partial x}$ represents the advection of a given field $\eta$ by $\psi$.

Equation~\ref{eq:WTG_PDE} needs boundary conditions, together with an expression of the forcing divergence, as done in the next section.  

\subsection{Definition of the domain and key parameters}

The domain considered is a channel along the equator with periodic boundary conditions in the zonal direction and no vortical flow along the meridional direction. The imposed boundary conditions are
\begin{equation}
\label{eq:boundary_conditions}
\frac{\partial \psi}{\partial y}|_{-L_y/2, L_y/2}=0, \,\, \psi(x,y,t)=\psi(x+L_x, y , t) 
\end{equation}
with $L_y=\pi L$ and $L_x=2 L_y/n$ with $n$ the aspect ratio of the domain. Here the domain will be fixed such that $L_y=4000$ km. This choice is made in order to encompass the dominant waves that are present in the Tropics \cite{Matsuno1966, Gill1980}.

The forcing term is expressed as,
\begin{eqnarray}
   \chi & = &A cos^2 (\pi y/L_y) cos (2 \pi x / L_x)+ B sin (\pi y/L_y) cos (2 \pi x / L_x) \label{eq:forcing} \\
   D & = & \nabla^2 \chi = - (2 \pi/L_x)^2 \left [ A cos^2 (\pi y/L_y) cos (2 \pi x / L_x)+ B sin (\pi y/L_y) cos (2 \pi x / L_x)  \right ] \nonumber \\ & + & 2 A (\pi/L_y)^2  \left [ sin^2(\pi y/L_y)-cos^2(\pi y /L_y) \right ] cos (2 \pi x / L_x) - B  (\pi/L_y)^2 sin (\pi y/L_y) cos (2 \pi x / L_x) \label{eq:divergence}
\end{eqnarray}
showing two different zones in the zonal direction, one being a source of divergence and a second being a sink. Moreover, the dependencies in the meridional direction characterize the presence of a symmetric source with a strong peak at the equator, and an asymmetric source. This idealized structure allows to clarify the impact of symmetries in the meridional direction. Other structures could be envisaged, which will be the subject of future works.

In principle, the parameters $A$ and $B$ are time dependent, but in the current analysis constant values are used. In the classical literature \citep{Trenberth2000, Kraucunas2007,Krishnamurti2013}, the estimation of the divergence field often varies between $10^{-6}$ and $10^{-5}$ s$^{-1}$. With this range and considering the maximum amplitude of the coefficients in \ref{eq:divergence}, one may estimate $A$ as
\begin{equation}
   A = \frac{D}{2} \left (\frac{L_y}{\pi} \right )^2 \approx [10^6,10^7] \,\, \rm{m^2/s} 
   \label{eq:estimation}
\end{equation}
A similar amplitude is assumed for parameter $B$. These amplitudes are compatible with the values of the velocity potential estimated from data \citep{Stanfield2024} and in comprehensive climate models \citep{Gastineau2009}. 

\subsection{The reduced order model}

Building reduced-order models for atmospheric flows has a long history back to the works of \cite{Saltzman1962}. The idea is to select a few equations describing a set of physical processes and to project them on key modes that will allow for a simplified description of the dynamics of interest \citep[e.g.][]{Veronis1973,Charney1980,Reinhold1982,Dijkstra2005,Pierini2011,Vannitsem2014,Vannitsem2015}. For the WTG model, 
the reduced-order formulation is obtained by projecting the equation~\ref{eq:WTG_PDE} on a set of Fourier modes as discussed in details in Appendix~\ref{sec:appA}. Using this procedure, a first reduced order model is built with 10 variables which is integrated using a second-order Heun method with a time step of 0.001 time unit, equivalent to about 30 seconds. A second model is developed with 36 variables in order to make a preliminary evaluation of the impact of the number of modes.

\section{Results}

We report here the computation of the Lyapunov exponents used to characterize the nature of the solutions. The definition and computation of the Lyapunov exponents, and related quantities are detailed in Appendix~\ref{sec:appB}.

\subsection{Chaos for n=0.20}

\begin{figure}
    \centering
    \includegraphics[width=0.8\linewidth]{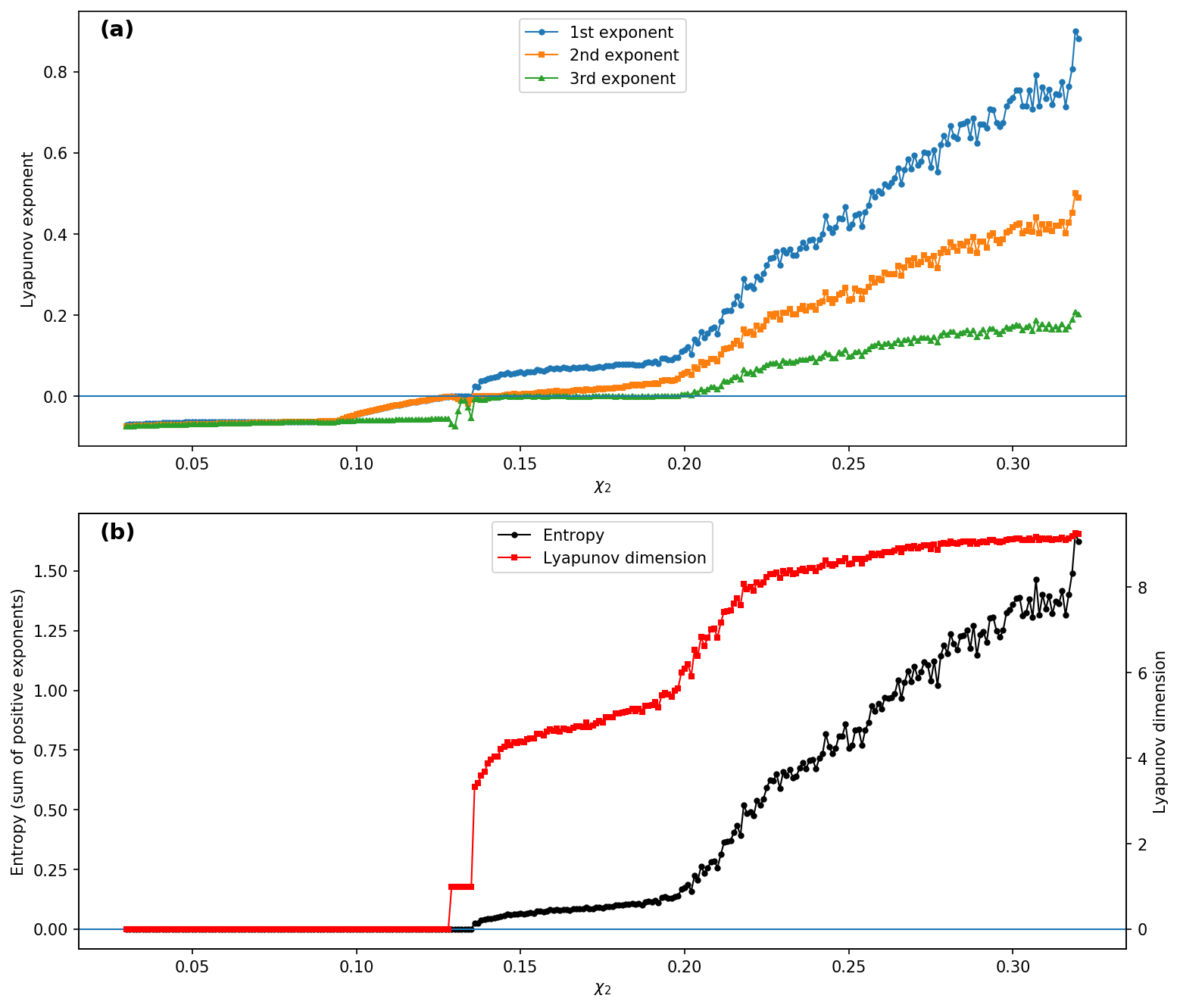}
    \caption{Lyapunov exponents (day$^{-1}$) for $n=0.20$ and a symmetric forcing around the equator. (a) The three first Lyapunov exponents as a function of the amplitude of the parameter $\chi_2$, (b) the Kolmogorov-Sinaï entropy and the Lyapunov dimension defined in Appendix~\ref{sec:appB}  as a function of $\chi_2$.}
    \label{fig:lyap-x1}
\end{figure}

Figure \ref{fig:lyap-x1}a displays the first three Lyapunov exponents as a function of the amplitude of the symmetric component\footnote{See Appendix~\ref{sec:appA} for more details about the forcing $\chi$ decomposition.} $\chi_2$ of $\chi$ for $n=0.20$. For small values of the forcing, a (stable) steady state is found. Once the amplitude of the forcing is increased, a succession of windows alternating periodic and chaotic solutions, is isolated. The first Lyapunov exponent can reach very large amplitudes for large forcing values. For still reasonable values of the forcing amplitude, between 0.1 and 0.2, the dominant Lyapunov exponent is low with values of less than 0.2 days$^{-1}$, suggesting a relatively high predictability in this range.

In Fig. \ref{fig:lyap-x1}b the Kolmogorov-Sinaï entropy and the Lyapunov dimension are displayed (defined in Appendix B). The striking feature of the entropy is to show a systematic increase, except in the windows of periodicity. At the same time the Lyapunov dimension increases considerably, even reaching values larger than 9 for a system limited to 10 dimensions. This peculiar feature contrasts with the experiments performed using extratropical models \citep{Vannitsem2015} for which the number of positive exponents are usually small. A possible reason is the multiplicative nature of the forcing in the WTG model. This question is worth addressing in future works.

To get some insights on the type of solutions that could emerge, several videos have been made (see \url{https://doi.org/10.5446/s_2048}). They show a dominant westward propagation of the large scale structures, with some rare eastward propagation. For large values of the forcing (beyond 0.2) the solutions look unrealistic as suggested by the very fast propagation of large-scale envelopes. 

\begin{figure}
    \centering
    \includegraphics[width=0.8\linewidth]{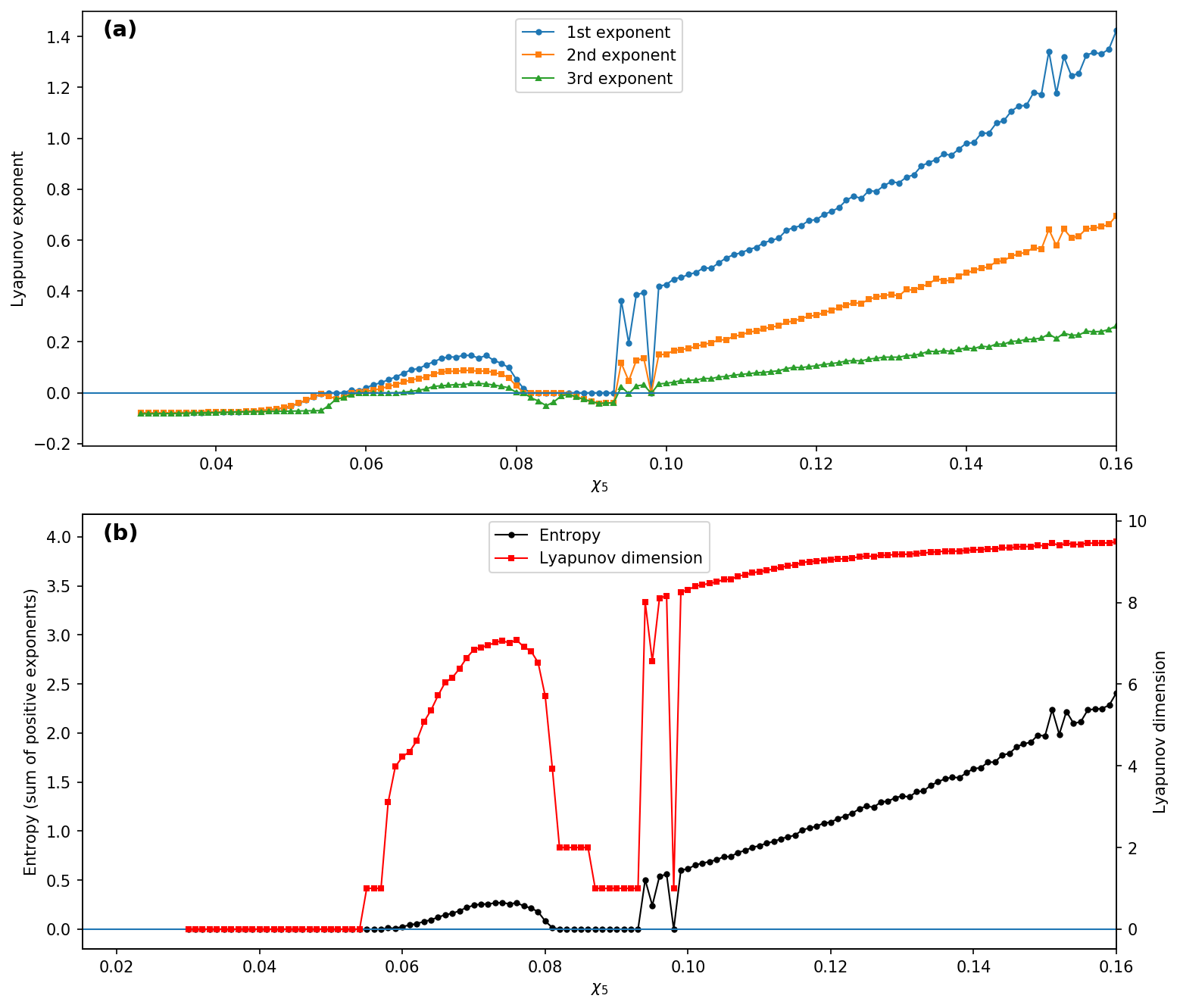}
     \caption{Lyapunov exponents (day$^{-1}$) for $n=0.20$ and an asymmetric forcing around the equator. (a) The three first Lyapunov exponents as a function of the amplitude of the parameter $\chi_5$, (b) the Kolmogorov-Sinaï entropy and the Lyapunov dimension defined in Appendix B as a function of $\chi_5$.}
    \label{fig:lyap-x4}
\end{figure}

Figure \ref{fig:lyap-x4}a displays the Lyapunov exponents when the forcing is purely asymmetric in the meridional direction. The emergence of chaos is here occurring sooner than for the symmetric forcing, starting at $\chi_5=0.065$, $\chi_5$ being the component controlling the antisymmetry of the forcing. The Lyapunov dimension also increases steadily together with the KS entropy (Fig. \ref{fig:lyap-x4}b). It is then followed by periodic and chaotic windows until about $0.11$. For larger values, chaos reemerges with very large Lyapunov exponents, Lyapunov dimensions and KS entropies. For even larger values of $\chi_5$, there is no solution anymore. 

\begin{figure}
    \centering
\includegraphics[width=0.8\linewidth]{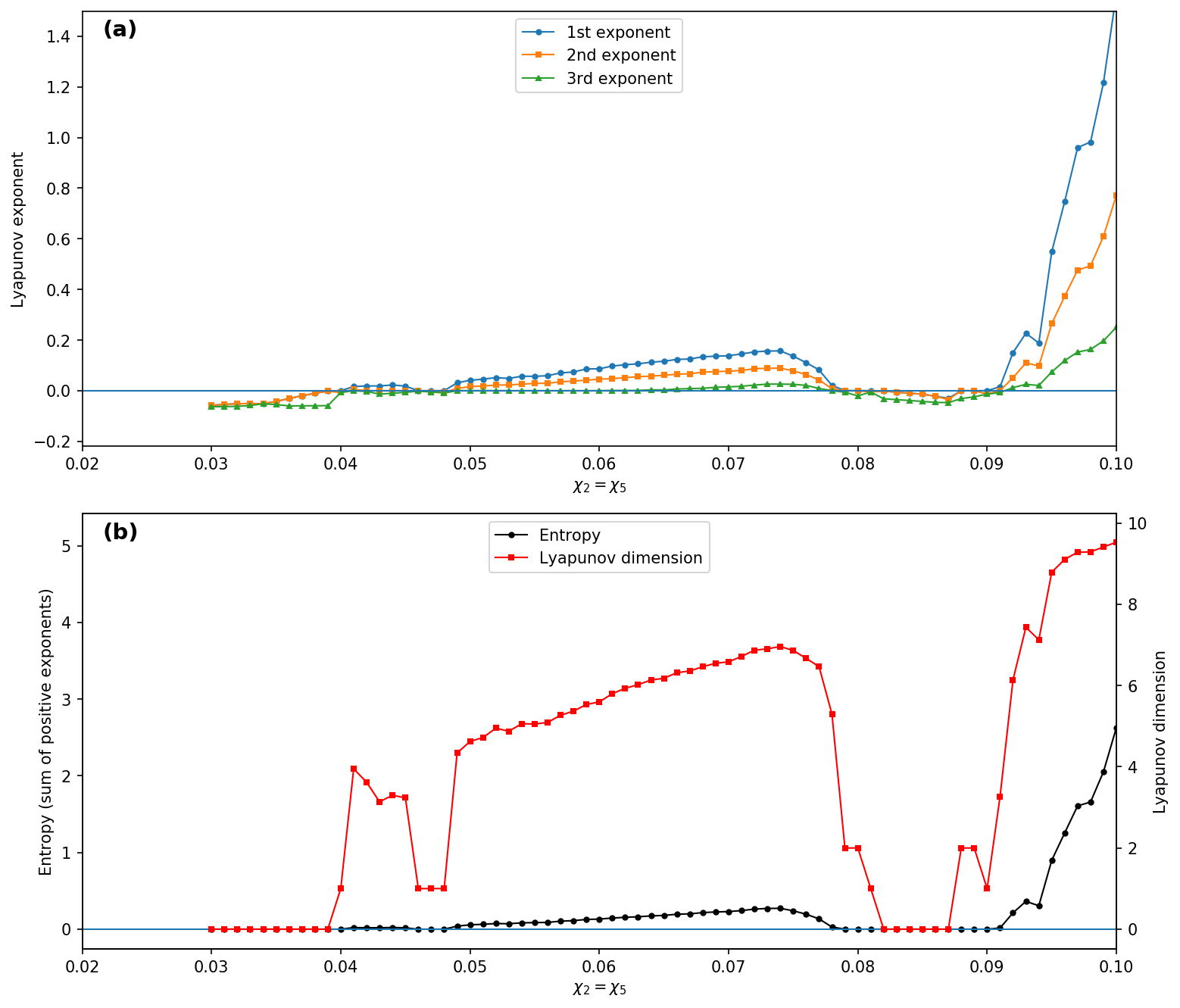}
\caption{Lyapunov exponents (day$^{-1}$) for $n=0.20$ and a combination of a symmetric and an asymmetric forcing around the equator. (a) The three first Lyapunov exponents as a function of the amplitude of the parameter $\chi_2=\chi_5$, (b) the Kolmogorov-Sinaï entropy and the Lyapunov dimension defined in Appendix B as a function of $\chi_2=\chi_5$.}
    \label{fig:lyap-x1x4}
\end{figure}

Finally, when both symmetric and anti-symmetric components of the forcing are present, the windows of alternating chaotic and periodic solutions are similar to the ones with the asymmetric forcing (Fig. \ref{fig:lyap-x1x4}). The chaotic behavior however starts for smaller amplitudes of each term of the forcing. Interestingly, the amplitudes of the exponents, the Lyapunov dimension and the entropy monotonically (almost linearly) increase in the window from about $\chi_2=\chi_5=0.05$ to $0.09$. For large values of the order of $0.1$, the instability is very large.

Movies are also provided (see \url{https://doi.org/10.5446/s_2048}) revealing similar conclusions to the symmetric case. For large values of the forcing, the propagation of the large-scale structures looks unrealistic.

In summary, chaos is emerging in the large-scale dynamics of the tropical atmosphere, whose nature (strong or weak) depends considerably on the spatial structure of the forcing. 

\subsection{Changing the aspect ratio, n}

\begin{figure}
    \centering
    \includegraphics[width=0.6\linewidth]{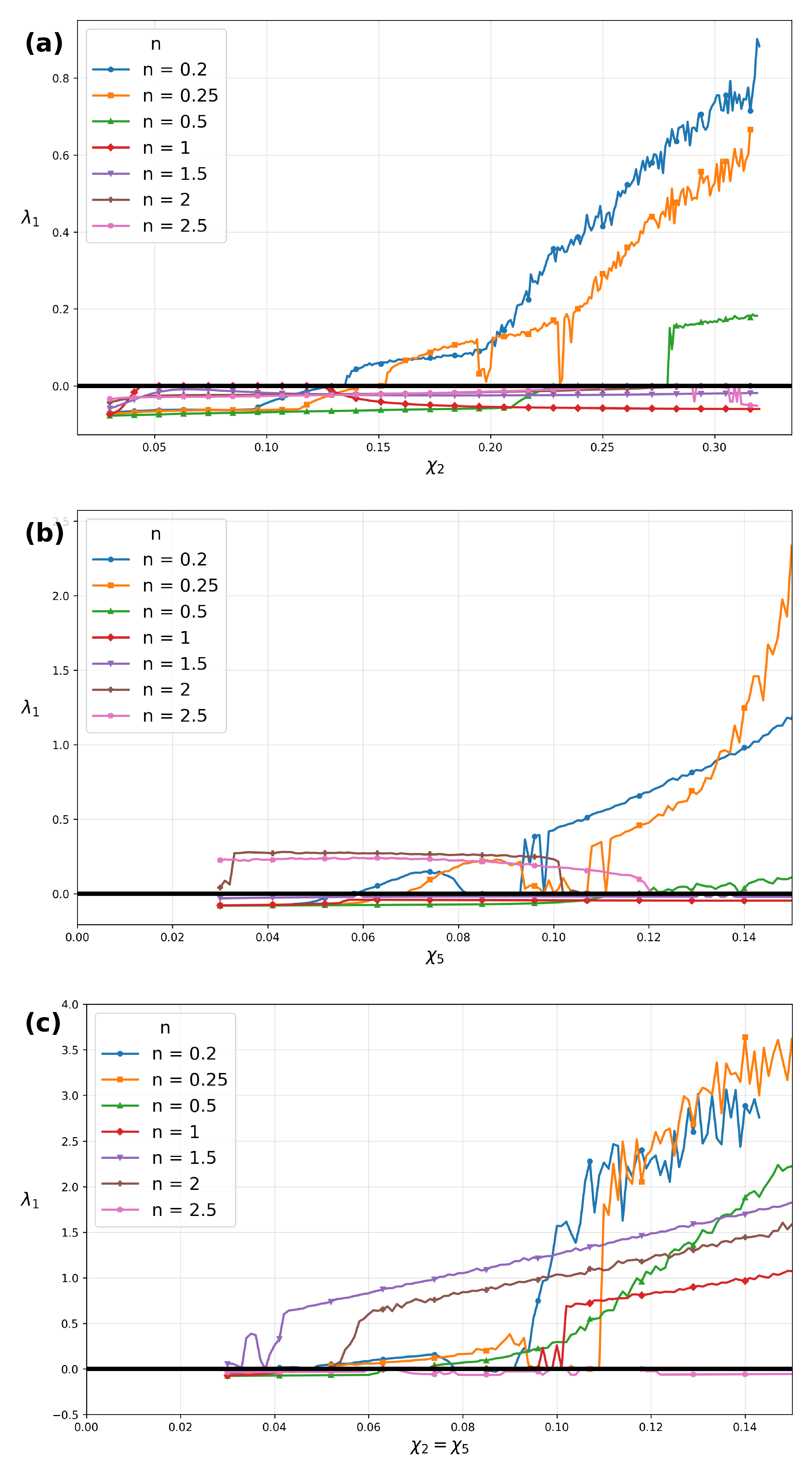}
    \caption{The first Lyapunov exponent as a function of the divergence forcing amplitude for different values of the aspect ratio $n$. (a) for the symmetric forcing, (b) for the asymmetric forcing, and (c) for the combination of the two types of forcing.}
    \label{fig:lyapvsn}
\end{figure}

Changing the aspect ratio, $n$, allows for investigating the interaction of the dynamical modes resolved by the model and the forcing at different longitudinal scales, the larger the value of $n$, the smaller the scales of the forcing and of the targeted modes. Figure \ref{fig:lyapvsn}a displays the first Lyapunov exponent as a function of $\chi_2$. For $n$ small (large domain), chaos emerges for large values of the forcing amplitude. Beyond $n$=1, the solutions are periodic or stationary, indicating that under symmetric forcing, emergence of chaos is unlikely at scales smaller than 10,000 km. The picture is different when considering asymmetric forcing as displayed in Figs. \ref{fig:lyapvsn}b and \ref{fig:lyapvsn}c where for large values of $n$, chaos is emerging for small values of the forcing amplitude. 

The results suggest that depending on the spatial structure of the forcing, chaos may emerge on various ranges of forcing amplitudes. Moreover, the emergence of chaos at planetary scales (from 20,000 to 40,000 km) is found for large values of the forcing, while it is occurring for small forcing values at scales of the order of a few thousands of kilometers. This has profound implications as scales of a few thousands of kilometers are first affected by the emergence of chaotic dynamics in the current system setting.

\subsection{Higher resolution model versions}

Galerkin truncation to a few number of modes could considerably affect the dynamics \citep{Reinhold1982}. In order to make a preliminary evaluation of such an impact, the number of modes is increased to 36 by using $M^{\max}=H^{\max}=P^{\max}=4$, with an aspect ratio of $n=0.20$.  

\begin{figure}
    \centering
    \includegraphics[width=0.8\linewidth]{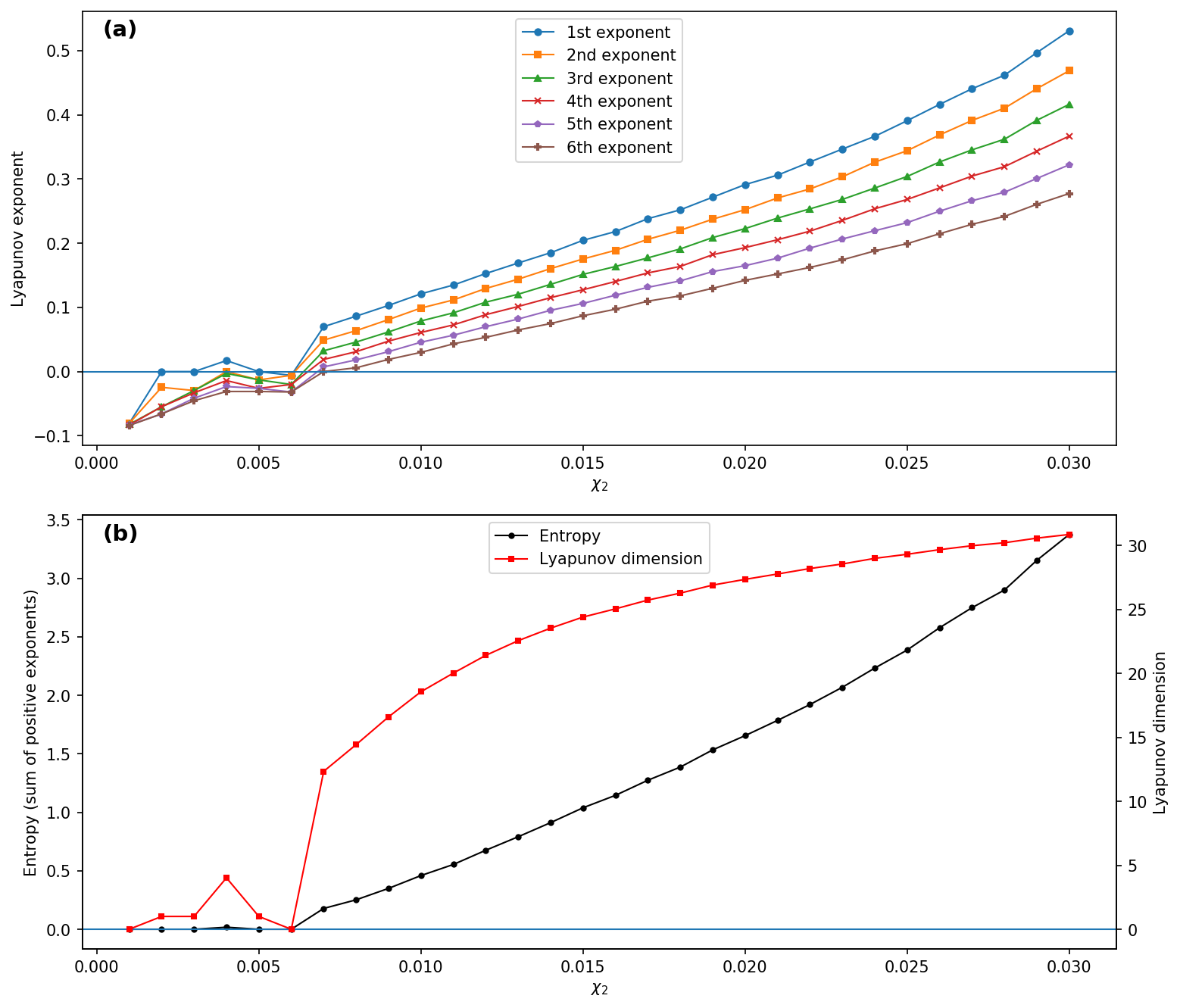}
    \caption{Lyapunov exponents for $n=0.20$ in the 36-dimensional model. The forcing $\chi$ is such that only the components $\chi_2$, $\chi_5$, $\chi_8$ and $\chi_{11}$ of this model are non-zero ($\chi_2 = \chi_5 = 5 \, \chi_8 = - \frac{5}{2} \, \chi_{11}$). Only the first six Lyapunov exponents as a function of the amplitude of the parameter $\chi_2$ are depicted.}
    \label{fig:lyap4x4}
\end{figure}

Figure \ref{fig:lyap4x4}a displays the first 6 exponents as a function of the forcing $\chi$ with $\chi_2 = \chi_5 = 5 \, \chi_8 = - \frac{5}{2} \, \chi_{11}$ being the non-zero components of its decomposition onto the model's modes.

Interestingly, the values of the positive exponents increase almost linearly with the forcing amplitude, and their number increases steadily. In panel \ref{fig:lyap4x4}b, the dimension and the entropy are also increasing steadily for forcing values larger than $0.006$. 

This experiment further support the extensive character of the instability properties of the reduced-order WTG model with the amplitude of the forcing. 

\section{Conclusions} 

Chaotic solutions are naturally emerging in reduced-order versions of the weak temperature gradient model describing the large-scale dynamics of the tropical atmosphere. This has been amply demonstrated in various versions of the reduced-order model with different parameter values for the divergence forcing, the scale of the tropical domain, and the spectral resolution of the model.

A notable result is the overall increase of the number of positive Lyapunov exponents, the Kolmogorov-Sinaï entropy and the Lyapunov dimension when the forcing amplitude is increased once chaos is settled. This feature contrasts with the dynamics found in other reduced-order models describing the dynamics of the atmosphere or the coupled ocean-atmosphere in the extratropics \cite{Vannitsem2015, Vannitsem2016, Vannitsem2017}. The origin of this difference is an important question to address in the future.

The spatial structure of the forcing is relatively arbitrary and other structures should be considered. A possible way is to project the effective divergence of the wind experienced in the Tropics on the modes of the model. This could be complemented with an analysis of the time dependence of the divergence and the emergence of pullback attractors in a more realistic setting \cite[e.g.][]{Ghil2008, Chekroun2011, Pierini2016, Vannitsem2021}.

Coming back to the question on the higher predictability of the atmosphere in the tropical regions alluded in the Introduction, our results suggest that the WTG tropical atmosphere could be highly sensitive to initial conditions depending on the choice of the amplitude of the forcing and the model's parameters. This feature depends on the setup of the model and in order to gain more insight into the matching between our results and the reality, we foresee two important lines of research to pursue: (i) adjusting the parameters of the model to the actual dynamics of the atmosphere, and (ii) clarifying the impact of incorporating other variables \citep[e.g.][]{Hottovy2015, Adames2022}, additional dissipation terms and/or surface interactions with land and ocean \cite[e.g.][]{Vannitsem2015, Xavier2024, Zhang2025}. These steps will be taken up in a near future. 

\begin{appendices}
\section{The wavenumber 2 truncated reduced-order tropical model equations}    
\label{sec:appA}

The full model equation~\ref{eq:WTG_PDE} is first non-dimensionalized, dividing the streamfunctions\footnote{From now on, the streamfunctions appearing in the equations are assumed to be the non-dimensional ones.} by a factor $L^2 / T$, with characteristic spatial scale $L = L_y / \pi$ and temporal scale $T = 1 / \sqrt{\beta c}$, with $c = \sqrt{g h_e}$ being the propagation velocity of the gravity waves in a shallow layer within the tropical atmosphere and $h_e$ the equivalent depth \citep{Kiladis2009}.
This yields the non-dimensional equation
\begin{equation}
    \label{eq:nondim_WTG_PDE}
    \frac{\partial}{\partial t}  \nabla^2 \psi = - J(\psi, \nabla^2 \psi) - \beta' \frac{\partial \psi}{\partial x} - \nabla \chi \cdot \nabla (\nabla^2 \psi + f') -  (\nabla^2 \psi + f') \nabla^2 \chi - r' \nabla^2 \psi
\end{equation}
with $r' = r T$, $\beta'=\beta L T$ and $f' = f'_0 + \beta' y$. 
The coordinates of the beta plane are also rescaled accordingly, i.e. being divided by $L$, leading to $x \in [0, 2\pi/n]$ and $y \in [-\pi/2, \pi/2]$. 

The non-dimensional forcing $\chi$ is provided by equation~\ref{eq:forcing} 
\begin{equation}
    \chi  = A' \cos^2 (y) \cos (n x)+ B' \sin (y) \cos (n x) \label{eq:forcing-nondim} 
\end{equation}
with $A = \frac{L^2}{T} A'$ and $B = \frac{L^2}{T} B'$. The nondimensionalized equation~\ref{eq:nondim_WTG_PDE} is then projected on Fourier modes satisfying the boundary conditions~\ref{eq:boundary_conditions}. Orthonormality, periodicity and no-flux boundary conditions lead to the following set of modes:
\begin{equation}
\label{eq:zonal_modes}
    F^A_P(x,y) = \left\{ \begin{aligned} 
  \sqrt{2} \ \sin(P \, y) & \quad \mathrm{if} \, P \, \mathrm{is \, odd} \\
  \sqrt{2} \ \cos(P \, y) & \quad \mathrm{if} \, P \, \mathrm{is \, even}
\end{aligned} \right.
\end{equation}
\begin{equation}
\label{eq:FKmodes}
    F^K_{M,P}(x,y) = \left\{ \begin{aligned} 
  2 \cos(M \, n  x) \, \cos(P \, y )&  \quad \mathrm{if} \, P \, \mathrm{is \, odd} \\
  2 \cos(M \, n  x) \, \sin(P \, y) & \quad \mathrm{if} \, P \, \mathrm{is \, even}
\end{aligned} \right.
\end{equation}
\begin{equation}
\label{eq:FLmodes}
    F^L_{H,P}(x,y) = \left\{ \begin{aligned} 
  2 \sin(H \, n  x) \, \cos(P \, y )&  \quad \mathrm{if} \, P \, \mathrm{is \, odd} \\
  2 \sin(H \, n  x) \, \sin(P \, y) & \quad \mathrm{if} \, P \, \mathrm{is \, even}
\end{aligned} \right.
\end{equation}
The alternation of sine and cosine of $y$ is due to the orthonormality condition with respect to the inner product
\begin{equation}
    \label{eq:inner_product}
    \langle S, G\rangle = \frac{n}{2\pi^2} \int_{-\pi/2}^{\pi/2} \int_0^{2\pi/n} S(x,y) \, F(x,y) \, \mathrm{d}x \, \mathrm{d}y
\end{equation}

It is worth noting that this condition allows for another valid choice for the purely zonal modes:
\begin{equation}
    \label{eq:other_zonal_modes}
    F^{A'}_P(x,y) = \left\{ \begin{aligned} 
  \sqrt{2} \ \cos(P \, y) & \quad \mathrm{if} \, P \, \mathrm{is \, odd} \\
  \sqrt{2} \ \sin(P \, y) & \quad \mathrm{if} \, P \, \mathrm{is \, even}
\end{aligned} \right.
\end{equation}

However, the advantage of using  the $F^A$ functions instead of the $F^{A'}$ ones is that $F^A_1 = \sqrt{2} \ \sin(y)$ allows for a representation of a global asymmetry of the basic fields in the model. Moreover, its derivative with respect to $y$ also allows for a direct representation of a symmetric zonal wind in the center of the domain. The modes given by equation~\ref{eq:zonal_modes} are therefore selected in the current work.

The reduced-order tropical models ordinary differential equations are obtained by projecting the equation~\ref{eq:nondim_WTG_PDE} onto a subset of the Fourier modes~\ref{eq:zonal_modes}, \ref{eq:FKmodes} and~\ref{eq:FLmodes}, using the inner product~\ref{eq:inner_product}. The modes are selected typically by truncating the wavenumbers to a certain level: $1 \leq P \leq P^{\max}$, $1 \leq M \leq M^{\max}$ and $1 \leq H \leq H^{\max}$. In the following, unless otherwise stated, and for the sake of simplicity; the modes will be reordered according to a single index, giving for a truncation at wavenumber $P^{\max}=M^{\max}=H^{\max}=2$:
\begin{equation}
\label{eq:modes}
\begin{aligned}
    F_1(x,y) & = \sqrt{2}\, \cos(y) \\
    F_2(x,y) & = 2 \cos(n x)\, \cos(y) \\
    F_3(x,y) & = 2 \sin(n x)\, \cos(y) \\
    F_4(x,y) & = \sqrt{2}\, \sin(2y) \\
    F_5(x,y) & = 2 \cos(n x)\, \sin(2y) \\
    F_6(x,y) & = 2 \sin(n x)\, \sin(2y) \\
    F_7(x,y) & = 2 \cos(2 n x) \, \cos(y) \\
    F_8(x,y) & = 2 \sin(2 n x) \, \cos(y) \\
    F_9(x,y) & = 2 \cos(2 n x) \, \sin(2y) \\
    F_{10}(x,y) & = 2 \sin(2 n x) \, \sin(2y)
\end{aligned}
\end{equation}

The non-dimensional forcing $\chi$ of the model given by equation~\ref{eq:forcing-nondim} is also projected on the modes~\ref{eq:modes}, giving
\begin{equation}
    \chi(x,y) = \chi_2 \, F_2(x,y) + \chi_5 \, F_5(x,y)
\end{equation}
with $\chi_2 = 4 A'/(3\pi)$ and $\chi_5 = 4 B'/(3\pi)$, providing the model internal representation of respectively the meridional symmetric and anti-symmetric components of the forcing.

The range of possible values for $\chi_2$ and $\chi_5$ can then be deduced from the estimations given in \ref{eq:estimation}, suggesting a typical range of $[0.01,0.1]$. In the current work, we sometimes explore values on a wider range up to $0.3$ that would correspond to situations for which strong divergence (corresponding to intense vertical transport) could be experienced. 

The actual projection of the equation~\ref{eq:nondim_WTG_PDE} onto the modes~\ref{eq:modes} is done using the LayerCake software~\citep{Demaeyer2026}, and results in the following set of general ordinary differential equations:
\begin{eqnarray}
\dot \psi_{1} & = & +\frac{32 \beta' \chi_4}{9 \pi} + (- r' - \sqrt{2} \chi_4 )\, \psi_{1}+\sqrt{2} \chi_5 \left(n^{2} + 1\right) \, \psi_{2}\nonumber \\ 
     & & +\sqrt{2} \chi_6 \left(n^{2} + 1\right) \, \psi_{3}+2 \sqrt{2} \chi_1 \, \psi_{4}- \frac{\sqrt{2} \chi_2 \left(n^{2} + 4\right)}{2} \, \psi_{5}\nonumber \\ 
     & & - \frac{\sqrt{2} \chi_3 \left(n^{2} + 4\right)}{2} \, \psi_{6}+\sqrt{2} \chi_9 \left(4 n^{2} + 1\right) \, \psi_{7}+\sqrt{2} \chi_{10} \left(4 n^{2} + 1\right) \, \psi_{8}\nonumber \\ 
     & & - 2 \sqrt{2} \chi_7 \left(n^{2} + 1\right) \, \psi_{9}- 2 \sqrt{2} \chi_8 \left(n^{2} + 1\right) \, \psi_{10} \\
\dot \psi_{2} & = & - \frac{\frac{16 \beta' \chi_5 \left(n^{2} + 4\right)}{9 \pi} - \frac{8 \beta' \chi_5}{3 \pi}}{n^{2} + 1} +\frac{\sqrt{2} \chi_5 \left(n^{2} + 2\right)}{2 \left(n^{2} + 1\right)} \, \psi_{1}+ (\frac{16 n^{2} \chi_7}{3 \pi} - r' + \sqrt{2} \chi_4 + \frac{4 \chi_7}{3 \pi} )\, \psi_{2}\nonumber \\ 
     & & +\frac{16 n^{4} \chi_8 + 20 n^{2} \chi_8 + 3 \pi n \beta' + 4 \chi_8}{3 \pi \left(n^{2} + 1\right)} \, \psi_{3}+\frac{2 \sqrt{2} \chi_2 \left(n^{2} - 1\right)}{n^{2} + 1} \, \psi_{4} \nonumber \\
     & & +\frac{\left(n^{2} + 4\right) \left(- 128 n^{2} \chi_9 - 15 \sqrt{2} \pi \chi_1 + 256 \chi_9 \left(n^{2} + 1\right) - 224 \chi_9\right)}{30 \pi \left(n^{2} + 1\right)} \, \psi_{5}\nonumber \\ 
     & & +\frac{16 \chi_{10} \left(n^{2} + 4\right) \left(4 n^{2} + 1\right)}{15 \pi \left(n^{2} + 1\right)} \, \psi_{6}+\frac{4 \chi_2 \left(- 8 n^{4} + 2 n^{2} + 1\right)}{3 \pi \left(n^{2} + 1\right)} \, \psi_{7}+\frac{4 \chi_3 \left(- 8 n^{4} + 2 n^{2} + 1\right)}{3 \pi \left(n^{2} + 1\right)} \, \psi_{8}\nonumber \\ 
     & & +\frac{64 \chi_5 \left(1 - 2 n^{2}\right)}{15 \pi} \, \psi_{9}+\frac{64 \chi_6 \left(1 - 2 n^{2}\right)}{15 \pi} \, \psi_{10}+\frac{8 \sqrt{2} n^{3}}{3 \pi \left(n^{2} + 1\right)} \, \psi_{1}\, \psi_{3}\nonumber \\ 
     & & - \frac{64 \sqrt{2} n^{3}}{15 \pi \left(n^{2} + 1\right)} \, \psi_{4}\, \psi_{6}+\frac{9 n \left(n^{2} - 1\right)}{2 \left(n^{2} + 1\right)} \, \psi_{5}\, \psi_{8}+\frac{9 n \left(1 - n^{2}\right)}{2 \left(n^{2} + 1\right)} \, \psi_{6}\, \psi_{7} \\
\dot \psi_{3} & = & - \frac{\frac{16 \beta' \chi_6 \left(n^{2} + 4\right)}{9 \pi} - \frac{8 \beta' \chi_6}{3 \pi}}{n^{2} + 1} +\frac{\sqrt{2} \chi_6 \left(n^{2} + 2\right)}{2 \left(n^{2} + 1\right)} \, \psi_{1}+\frac{16 n^{4} \chi_8 + 20 n^{2} \chi_8 - 3 \pi n \beta' + 4 \chi_8}{3 \pi \left(n^{2} + 1\right)} \, \psi_{2}\nonumber \\ 
     & & + (- \frac{16 n^{2} \chi_7}{3 \pi} - r' + \sqrt{2} \chi_4 - \frac{4 \chi_7}{3 \pi} )\, \psi_{3}+\frac{2 \sqrt{2} \chi_3 \left(n^{2} - 1\right)}{n^{2} + 1} \, \psi_{4}+\frac{16 \chi_{10} \left(n^{2} + 4\right) \left(4 n^{2} + 1\right)}{15 \pi \left(n^{2} + 1\right)} \, \psi_{5}\nonumber \\ 
     & & +\frac{\left(n^{2} + 4\right) \left(128 n^{2} \chi_9 - 15 \sqrt{2} \pi \chi_1 - 256 \chi_9 \left(n^{2} + 1\right) + 224 \chi_9\right)}{30 \pi \left(n^{2} + 1\right)} \, \psi_{6}\nonumber \\
     & & +\frac{4 \chi_3 \left(8 n^{4} - 2 n^{2} - 1\right)}{3 \pi \left(n^{2} + 1\right)} \, \psi_{7}+\frac{4 \chi_2 \left(- 8 n^{4} + 2 n^{2} + 1\right)}{3 \pi \left(n^{2} + 1\right)} \, \psi_{8}\nonumber \\ 
     & & +\frac{64 \chi_6 \left(2 n^{2} - 1\right)}{15 \pi} \, \psi_{9}+\frac{64 \chi_5 \left(1 - 2 n^{2}\right)}{15 \pi} \, \psi_{10}- \frac{8 \sqrt{2} n^{3}}{3 \pi \left(n^{2} + 1\right)} \, \psi_{1}\, \psi_{2}\nonumber \\ 
     & & +\frac{64 \sqrt{2} n^{3}}{15 \pi \left(n^{2} + 1\right)} \, \psi_{4}\, \psi_{5}+\frac{9 n \left(1 - n^{2}\right)}{2 \left(n^{2} + 1\right)} \, \psi_{5}\, \psi_{7}+\frac{9 n \left(1 - n^{2}\right)}{2 \left(n^{2} + 1\right)} \, \psi_{6}\, \psi_{8}
\end{eqnarray}
\newpage
\begin{eqnarray}
\dot \psi_{4} & = & +\frac{8 \beta' \chi_1}{9 \pi} - \frac{\sqrt{2} \chi_1}{4} \, \psi_{1}+\frac{\sqrt{2} \chi_2 \left(n^{2} + 1\right)}{4} \, \psi_{2}\nonumber \\ 
     & & +\frac{\sqrt{2} \chi_3 \left(n^{2} + 1\right)}{4} \, \psi_{3}- r' \, \psi_{4}+\frac{\sqrt{2} \chi_7 \left(4 n^{2} + 1\right)}{4} \, \psi_{7}\nonumber \\ 
     & & +\frac{\sqrt{2} \chi_8 \left(4 n^{2} + 1\right)}{4} \, \psi_{8}+\frac{16 \sqrt{2} n}{5 \pi} \, \psi_{2}\, \psi_{6}- \frac{16 \sqrt{2} n}{5 \pi} \, \psi_{3}\, \psi_{5}\nonumber \\ 
     & & +\frac{32 \sqrt{2} n}{5 \pi} \, \psi_{7}\, \psi_{10}- \frac{32 \sqrt{2} n}{5 \pi} \, \psi_{8}\, \psi_{9} \\
\dot \psi_{5} & = & - \frac{\frac{16 \beta' \chi_2 \left(n^{2} + 1\right)}{9 \pi} + \frac{8 \beta' \chi_2}{3 \pi}}{n^{2} + 4} +\frac{\sqrt{2} \chi_2 \left(n^{2} + 2\right)}{2 \left(n^{2} + 4\right)} \, \psi_{1}+\frac{\left(n^{2} + 1\right) \left(- 64 n^{2} \chi_9 + 15 \sqrt{2} \pi \chi_1 + 128 \chi_9 \left(n^{2} + 1\right) - 16 \chi_9\right)}{15 \pi \left(n^{2} + 4\right)} \, \psi_{2}\nonumber \\ 
     & & +\frac{16 \chi_{10} \left(n^{2} + 1\right) \left(4 n^{2} + 7\right)}{15 \pi \left(n^{2} + 4\right)} \, \psi_{3}+\frac{64 n^{2} \chi_7 - 15 \pi r' + 16 \chi_7}{15 \pi} \, \psi_{5}+\frac{64 n^{4} \chi_8 + 272 n^{2} \chi_8 + 15 \pi n \beta' + 64 \chi_8}{15 \pi \left(n^{2} + 4\right)} \, \psi_{6}\nonumber \\ 
     & & +\frac{16 \chi_5 \left(- 8 n^{4} + 26 n^{2} + 7\right)}{15 \pi \left(n^{2} + 4\right)} \, \psi_{7}+\frac{16 \chi_6 \left(- 8 n^{4} + 26 n^{2} + 7\right)}{15 \pi \left(n^{2} + 4\right)} \, \psi_{8}+\frac{64 \chi_2 \left(- 2 n^{4} - n^{2} + 1\right)}{15 \pi \left(n^{2} + 4\right)} \, \psi_{9}\nonumber \\ 
     & & +\frac{64 \chi_3 \left(- 2 n^{4} - n^{2} + 1\right)}{15 \pi \left(n^{2} + 4\right)} \, \psi_{10}+\frac{32 \sqrt{2} n \left(n^{2} + 3\right)}{15 \pi \left(n^{2} + 4\right)} \, \psi_{1}\, \psi_{6}- \frac{9 n^{3}}{2 n^{2} + 8} \, \psi_{2}\, \psi_{8}\nonumber \\ 
     & & +\frac{64 \sqrt{2} n \left(3 - n^{2}\right)}{15 \pi \left(n^{2} + 4\right)} \, \psi_{3}\, \psi_{4}+\frac{9 n^{3}}{2 \left(n^{2} + 4\right)} \, \psi_{3}\, \psi_{7} \\
\dot \psi_{6} & = & - \frac{\frac{16 \beta' \chi_3 \left(n^{2} + 1\right)}{9 \pi} + \frac{8 \beta' \chi_3}{3 \pi}}{n^{2} + 4} +\frac{\sqrt{2} \chi_3 \left(n^{2} + 2\right)}{2 \left(n^{2} + 4\right)} \, \psi_{1}+\frac{16 \chi_{10} \left(n^{2} + 1\right) \left(4 n^{2} + 7\right)}{15 \pi \left(n^{2} + 4\right)} \, \psi_{2}\nonumber \\ 
     & & +\frac{\left(n^{2} + 1\right) \left(64 n^{2} \chi_9 + 15 \sqrt{2} \pi \chi_1 - 128 \chi_9 \left(n^{2} + 1\right) + 16 \chi_9\right)}{15 \pi \left(n^{2} + 4\right)} \, \psi_{3}+\frac{64 n^{4} \chi_8 + 272 n^{2} \chi_8 - 15 \pi n \beta' + 64 \chi_8}{15 \pi \left(n^{2} + 4\right)} \, \psi_{5}\nonumber \\ 
     & & +\frac{16 \chi_6 \left(8 n^{4} - 26 n^{2} - 7\right)}{15 \pi \left(n^{2} + 4\right)} \, \psi_{7}+\frac{16 \chi_5 \left(- 8 n^{4} + 26 n^{2} + 7\right)}{15 \pi \left(n^{2} + 4\right)} \, \psi_{8}+\frac{64 \chi_3 \left(2 n^{4} + n^{2} - 1\right)}{15 \pi \left(n^{2} + 4\right)} \, \psi_{9}\nonumber \\ 
     & & +\frac{64 \chi_2 \left(- 2 n^{4} - n^{2} + 1\right)}{15 \pi \left(n^{2} + 4\right)} \, \psi_{10}+\frac{32 \sqrt{2} n \left(- n^{2} - 3\right)}{15 \pi \left(n^{2} + 4\right)} \, \psi_{1}\, \psi_{5}+\frac{64 \sqrt{2} n \left(n^{2} - 3\right)}{15 \pi \left(n^{2} + 4\right)} \, \psi_{2}\, \psi_{4}\nonumber \\ 
     & & +\frac{9 n^{3}}{2 \left(n^{2} + 4\right)} \, \psi_{2}\, \psi_{7}+\frac{9 n^{3}}{2 \left(n^{2} + 4\right)} \, \psi_{3}\, \psi_{8} +\frac{- 64 n^{2} \chi_7 - 15 \pi r' - 16 \chi_7}{15 \pi} \, \psi_{6} \\
\dot \psi_{7} & = & - \frac{\frac{64 \beta' \chi_9 \left(n^{2} + 1\right)}{9 \pi} - \frac{8 \beta' \chi_9}{3 \pi}}{4 n^{2} + 1} +\frac{\sqrt{2} \chi_9 \left(2 n^{2} + 1\right)}{4 n^{2} + 1} \, \psi_{1}+\frac{4 \chi_2 \left(n^{2} + 1\right)}{3 \pi} \, \psi_{2}\nonumber \\ 
     & & +\frac{4 \chi_3 \left(- n^{2} - 1\right)}{3 \pi} \, \psi_{3}+\frac{2 \sqrt{2} \chi_7 \left(4 n^{2} - 1\right)}{4 n^{2} + 1} \, \psi_{4}+\frac{16 \chi_5 \left(n^{2} + 4\right)}{15 \pi} \, \psi_{5}\nonumber \\ 
     & & +\frac{16 \chi_6 \left(- n^{2} - 4\right)}{15 \pi} \, \psi_{6}+ (- r' + \sqrt{2} \chi_4 )\, \psi_{7}+\frac{2 n \beta'}{4 n^{2} + 1} \, \psi_{8}\nonumber \\ 
     & & - \frac{2 \sqrt{2} \chi_1 \left(n^{2} + 1\right)}{4 n^{2} + 1} \, \psi_{9}+\frac{64 \sqrt{2} n^{3}}{3 \pi \left(4 n^{2} + 1\right)} \, \psi_{1}\, \psi_{8}- \frac{9 n}{8 n^{2} + 2} \, \psi_{2}\, \psi_{6}\nonumber \\ 
     & & - \frac{9 n}{8 n^{2} + 2} \, \psi_{3}\, \psi_{5}- \frac{512 \sqrt{2} n^{3}}{15 \pi \left(4 n^{2} + 1\right)} \, \psi_{4}\, \psi_{10}
\end{eqnarray}
\newpage
\begin{eqnarray}
\dot \psi_{8} & = & - \frac{\frac{64 \beta' \chi_{10} \left(n^{2} + 1\right)}{9 \pi} - \frac{8 \beta' \chi_{10}}{3 \pi}}{4 n^{2} + 1} +\frac{\sqrt{2} \chi_{10} \left(2 n^{2} + 1\right)}{4 n^{2} + 1} \, \psi_{1}+\frac{4 \chi_3 \left(n^{2} + 1\right)}{3 \pi} \, \psi_{2}\nonumber \\ 
     & & +\frac{4 \chi_2 \left(n^{2} + 1\right)}{3 \pi} \, \psi_{3}+\frac{2 \sqrt{2} \chi_8 \left(4 n^{2} - 1\right)}{4 n^{2} + 1} \, \psi_{4}+\frac{16 \chi_6 \left(n^{2} + 4\right)}{15 \pi} \, \psi_{5}\nonumber \\ 
     & & +\frac{16 \chi_5 \left(n^{2} + 4\right)}{15 \pi} \, \psi_{6}- \frac{2 n \beta'}{4 n^{2} + 1} \, \psi_{7}+ (- r' + \sqrt{2} \chi_4 )\, \psi_{8}\nonumber \\ 
     & & - \frac{2 \sqrt{2} \chi_1 \left(n^{2} + 1\right)}{4 n^{2} + 1} \, \psi_{10}- \frac{64 \sqrt{2} n^{3}}{3 \pi \left(4 n^{2} + 1\right)} \, \psi_{1}\, \psi_{7}+\frac{9 n}{2 \left(4 n^{2} + 1\right)} \, \psi_{2}\, \psi_{5}\nonumber \\ 
     & & - \frac{9 n}{8 n^{2} + 2} \, \psi_{3}\, \psi_{6}+\frac{512 \sqrt{2} n^{3}}{15 \pi \left(4 n^{2} + 1\right)} \, \psi_{4}\, \psi_{9} \\
\dot \psi_{9} & = & - \frac{\frac{16 \beta' \chi_7 \left(4 n^{2} + 1\right)}{9 \pi} + \frac{8 \beta' \chi_7}{3 \pi}}{4 n^{2} + 4} +\frac{\sqrt{2} \chi_7 \left(2 n^{2} + 1\right)}{4 \left(n^{2} + 1\right)} \, \psi_{1}+\frac{4 \chi_5 \left(4 n^{2} + 7\right)}{15 \pi} \, \psi_{2}\nonumber \\ 
     & & +\frac{4 \chi_6 \left(- 4 n^{2} - 7\right)}{15 \pi} \, \psi_{3}+\frac{4 \chi_2 \left(n^{2} + 4\right) \left(4 n^{2} + 1\right)}{15 \pi \left(n^{2} + 1\right)} \, \psi_{5}+\frac{4 \chi_3 \left(- 4 n^{2} - 1\right) \left(n^{2} + 4\right)}{15 \pi \left(n^{2} + 1\right)} \, \psi_{6}\nonumber \\ 
     & & +\frac{\sqrt{2} \chi_1 \left(4 n^{2} + 1\right)}{4 \left(n^{2} + 1\right)} \, \psi_{7}- \frac{4 r' \left(n^{2} + 1\right)}{4 n^{2} + 4} \, \psi_{9}+\frac{2 n \beta'}{4 n^{2} + 4} \, \psi_{10}\nonumber \\ 
     & & +\frac{16 \sqrt{2} n \left(4 n^{2} + 3\right)}{15 \pi \left(n^{2} + 1\right)} \, \psi_{1}\, \psi_{10}+\frac{32 \sqrt{2} n \left(3 - 4 n^{2}\right)}{15 \pi \left(n^{2} + 1\right)} \, \psi_{4}\, \psi_{8} \\
\dot \psi_{10} & = & - \frac{\frac{16 \beta' \chi_8 \left(4 n^{2} + 1\right)}{9 \pi} + \frac{8 \beta' \chi_8}{3 \pi}}{4 n^{2} + 4} +\frac{\sqrt{2} \chi_8 \left(2 n^{2} + 1\right)}{4 \left(n^{2} + 1\right)} \, \psi_{1}+\frac{4 \chi_6 \left(4 n^{2} + 7\right)}{15 \pi} \, \psi_{2}\nonumber \\ 
     & & +\frac{4 \chi_5 \left(4 n^{2} + 7\right)}{15 \pi} \, \psi_{3}+\frac{4 \chi_3 \left(n^{2} + 4\right) \left(4 n^{2} + 1\right)}{15 \pi \left(n^{2} + 1\right)} \, \psi_{5}+\frac{4 \chi_2 \left(n^{2} + 4\right) \left(4 n^{2} + 1\right)}{15 \pi \left(n^{2} + 1\right)} \, \psi_{6}\nonumber \\ 
     & & +\frac{\sqrt{2} \chi_1 \left(4 n^{2} + 1\right)}{4 \left(n^{2} + 1\right)} \, \psi_{8}- \frac{2 n \beta'}{4 n^{2} + 4} \, \psi_{9}- \frac{4 r' \left(n^{2} + 1\right)}{4 n^{2} + 4} \, \psi_{10}\nonumber \\ 
     & & +\frac{16 \sqrt{2} n \left(- 4 n^{2} - 3\right)}{15 \pi \left(n^{2} + 1\right)} \, \psi_{1}\, \psi_{9}+\frac{32 \sqrt{2} n \left(4 n^{2} - 3\right)}{15 \pi \left(n^{2} + 1\right)} \, \psi_{4}\, \psi_{7}
\end{eqnarray}

These equations are integrated in time using a second order Heun integrator with an non-dimensional time step of $0.001$, corresponding to a dimensional time step of about 30 seconds. Note that a certain number of terms are disappearing depending on the projection of the forcing $\chi$ on the set of modes.

A second version of the model with a truncation up to wavenumber 4 ($M^{\max}=H^{\max}=P^{\max}=4$) is also developed with the same procedure and tested. It leads to a set of 36 variables whose equations can be obtained using the Python code provided as supplementary material.

The forcing is now projected on this set of modes as follows,
\begin{equation}
 \chi(x,y) = \chi_2 \, F_2(x,y) + \chi_5 \, F_5(x,y) + \chi_8 \, F_8(x,y) + \chi_{11} \, F_{11}(x,y)
\end{equation}
 where the modes $F_2$ and $F_5$ expressions are provided by~(\ref{eq:modes}), but the modes $F_8$ and $F_{11}$ on the other hand correspond respectively in this configuration to the functions $2 \cos(3y)\cos(n x)$ and $2 \sin(4y) \cos(nx)$. In term of the forcing $\chi$ given by equation~(\ref{eq:forcing-nondim}), its projection on the modes gives the following potentially non-zero coefficients: $\chi_2 = 4 A'/(3\pi)$, $\chi_5 = 4 B'/(3\pi)$, $\chi_8 = 4 A'/(15\pi)$ and $\chi_{11} = - 8 B'/(15\pi)$.
 $\chi_2$ and $\chi_8$ provide the model internal representation of the meridional symmetric component of the forcing, while $\chi_5$ and $\chi_{11}$ provide the anti-symmetric one.

\section{The Lyapunov exponents}    
\label{sec:appB}

A nice description of the definitions and properties of the exponents is provided in \cite{Eckmann1985}. The Lyapunov exponents characterize the exponential divergences of infinitesimally small initial errors. If the dynamical system under consideration is described by $M$ ordinary differential equations, this system possesses $M$ Lyapunov exponents, referred to as the {\it Lyapunov spectrum}. If at least one of them is positive, the solutions generated by the system show the exponential divergence of infinitesimally small perturbations. If the largest one is zero, the solutions are either periodic or quasi-periodic. If all negative, the solution is a steady state (or a fixed point). For a detailed description of the theory and the algorithms to compute the exponents \citep{Eckmann1985, Parker1989, Kuptsov2012, Frederiksen2023}. For specific implementations in reduced-order or intermediate complexity models, see for instance \citep{Legras1985, Vannitsem1997, Vannitsem2016, DeCruz2018}.

To define the exponents, let us consider a dynamical system in the form
\begin{equation}
\frac{d\bf{x}}{dt} = {\bf f}(\bf x, \lambda)
\label{equat}
\end{equation}
where $\bf{x}$ is a vector of relevant variables
$\bf x$ = $( x_1, ..., x_N)$ and $t$ the time.
The vector $\bf f$ summarizes the impact of the dynamical processes responsible for the change of $\bf x$, and
$\lambda$ denotes all the parameters present in the description of the dynamics.

The formal solution of Eq. (\ref{equat}) is
\begin{equation}
\bf x (t) = G_t (\bf x_0, \lambda)
\label{traject}
\end{equation}
whith the initial state ${\bf x}(t_0) = {\bf x}_0$. The operator $G_t$ is the resolvent.

Let us now consider two initial states separated by en error, $\delta \bf x_0 =\bf x'_0-\bf x_0$. The perturbed initial state generates a new trajectory in phase space which results in an error with the reference trajectory, denoted as $\delta \bf x (t)$. Provided that this perturbation is sufficiently small, its dynamics is governed by the linearized equation,
\begin{equation}
\frac{d\delta \bf x}{dt} =  \frac{\partial \bf f}{\partial \bf x}_{\vert \bf {x}(t)}
\delta \bf{x}
\label {linear}
\end{equation}
whose formal solution is
\begin{equation}
\delta \bf x (t) = {\bf M}(t,\bf{x}(t_0)) \delta \bf{x} (t_0)
\end{equation}

For infinitesimally small initial errors and provided the system is ergodic, the divergence of initially closed states for infinite positive times is determined by the logarithm of the eigenvalues of the
matrix $(\bf{M}^T \bf{M})^{2(t-t_0)}$, referred to as the Lyapunov exponents and denoted as $\lambda_i$ for $i=1,..., N$. The same exponents are obtained when considering the matrix $(\bf{M} \bf{M}^T)^{2(t-t_0)}$ for $t_0$ going to -$\infty$.

Several methods have been developed to numerically evaluate the Lyapunov exponents \citep{Parker1989,Kuptsov2012}.
One of the most popular method consists in following the evolution of a set of
orthonormal vectors chosen initially at random in the tangent space of the trajectory $\bf x(t)$. These vectors are evolved using the linearized evolution equations (Eq. \ref{linear}) and regularly orthonormalized using the Gram-Schmidt method. After a rapid transient, the first vector of this set, free of any constraint, will tend to the direction of maximal stretching associated to the largest Lyapunov exponent; the second vector, orthogonal to the previous one, will tend to the second most unstable direction; and
so on. The total amplification along these different vectors in the tangent space will provide the associated Lyapunov exponents.

From these exponents, two quantities that will be used in the present paper, can be defined: The Kolmogorov-Sinaï entropy (KSE) and the Lyapunov dimension ($D_L$). The former is given by
\begin{equation}
    KSE= \sum_{i=1}^{I} \lambda_i
\end{equation}
where $I$ is the number of positive exponents. It characterizes the volume amplification in the subspace of the unstable directions. The latter is given by
\begin{equation}
D_L = j + \frac{\sum_{i=1}^{j}\lambda_i} {\left|\lambda_{j+1}\right|},
\end{equation}
with $\lambda_{j+1}$ the first exponent below zero. This quantity characterizes the dimension of the volume preserved by the flow for infinite time. This dimension is usually conjectured to be close to the (fractal when appropriate) dimension of the underlying attractor.

\section{Code availability}
The code is available on Github at \url{https://github.com/Climdyn/WTG-TM}. Additionally, it is archived on Zenodo~\citep{demaeyer_wtg_2026}.

\section{Video supplements}
The time evolution of the model dynamics is illustrated by a set of videos, which are available
online\\ at \href{https://doi.org/10.5446/s_2048}{doi.org:10.5446/s\_2048}. 
\end{appendices}

\bibliographystyle{unsrt}  
\bibliography{file}  

\end{document}